%% file: main.tex
\documentclass[12pt]{amsart}
\usepackage{enumerate}
\usepackage{enumitem}
\usepackage[colorlinks=true, linkcolor=blue, urlcolor=blue, citecolor=blue, anchorcolor=blue, pdfborder={0 0 0}]{hyperref}
\usepackage{url}
\usepackage{graphicx,color,colortbl}
\usepackage[dvipsnames,table,xcdraw]{xcolor}
\usepackage{cite}
\usepackage{amsthm, amsmath, amssymb}
\usepackage{mathtools}
\usepackage[top=45truemm, bottom=45truemm, left=30truemm, right=30truemm]{geometry}
\usepackage{subcaption}
\usepackage{cancel}
\usepackage{float}
\usepackage{tabularx}
\usepackage{makecell}
\usepackage{array}
\usepackage{ragged2e}
\usepackage{csquotes}
\usepackage{booktabs}

\usepackage{tikz}
\usetikzlibrary{arrows,positioning,calc,shapes}
\usepackage{graphicx}
\usepackage{mdframed}
\usepackage{mathtools}
\usepackage{algorithm2e}
\usepackage{arydshln}

\usepackage[toc,acronym,symbols]{glossaries}

\makenoidxglossaries

\usepackage{epigraph}
\loadglsentries{gloss}

\usepackage{amsthm}
\usepackage{thmtools}

\newtheorem{definition}{Definition}[section]

\newtheorem{lemma}{Lemma}[section]

\theoremstyle{definition}
\newtheorem{remark}{Remark}[section]
\newtheorem{example}{Example}[section]
\newtheorem{construction}{Construction}[section]

\newcommand{\define}{\overset{\text{\normalfont\fontsize{5}{6}\selectfont def}}{=}}
\newcommand{\samp}{\xleftarrow{\text{\normalfont\fontsize{5}{6}\selectfont\$}}}

\DeclareMathOperator*{\argmin}{arg\,min}
\DeclareMathOperator*{\Ex}{E}

\title[Biometric System Constructions]{A Framework for the Security and Privacy of Biometric System Constructions under Defined Computational Assumptions}

\author[Sam Grierson]{\hspace{1mm}Sam Grierson}
\address{Sam Grierson\\Edinburgh Napier University\\Edinburgh, UK}
\email{s.grierson2@napier.ac.uk}

\author[William J Buchanan]{\href{https://orcid.org/0000-0003-0809-3523}{\includegraphics[scale=0.06]{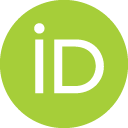}}\hspace{1mm}William J Buchanan}
\address{William J Buchanan\\Blockpass ID Lab\\Edinburgh Napier University\\Edinburgh, UK}
\email{b.buchanan@napier.ac.uk}

\author[Craig Thomson]{\hspace{1mm}Craig Thomson}
\address{Craig Thomson\\Blockpass ID Lab\\Edinburgh Napier University, UK}

\author[Baraq Galeb]{\hspace{1mm}Baraq Galeb}
\address{Baraq Galeb\\Blockpass ID Lab\\Edinburgh Napier University\\Edinburgh, UK}

\author[Chris Eckl]{\hspace{1mm}Chris Eckl}
\address{Chris Eckl\\Condastis\\Edinburgh, UK}
\email{chris.eckl@condatis.com}

\begin{document}

\begingroup
\let\MakeUppercase\relax
\maketitle
\endgroup
\begin{abstract}
Biometric systems, while offering convenient authentication, often fall short in providing rigorous security assurances. A primary reason is the ad-hoc design of protocols and components, which hinders the establishment of comprehensive security proofs. This paper introduces a formal framework for constructing secure and privacy-preserving biometric systems. By leveraging the principles of universal composability, we enable the modular analysis and verification of individual system components. This approach allows us to derive strong security and privacy properties for the entire system, grounded in well-defined computational assumptions.

\end{abstract}

\section{Introduction}\label{ch:2}
\epigraph{%
  \itshape Staying alive is not enough to guarantee survival. Development is the best way to ensure
  survival.
}{%
  --- Liu Cixin, \textit{The Dark Forest}, 2008
}

\noindent  Biometrics are now a common mode of authentication for end users,
particularly those who use modern smartphones; for example, a Cisco Duo report surveying 50 million devices globally found that 81\% of smartphones had biometrics enabled \cite{Lew22}. The uptake in using biometrics reflects the recent advancements in the practicality user focused biometric authentication methods \cite{Bud18,RE20}, such as Apple's FaceID \cite{App22} and  Android-based \cite{Bha+15}. Recently, the paradigm in society has shifted toward a more privacy-conscious users \cite{Zub15,Zub19}. The consequence of this paradigm shift is a change in the user's perception of the security and confidentiality of using their biometrics for authentication, primarily influenced by the number of cases where their biometric data was used without consent
\cite{Mec+18, ITL20}.

The societal shift in focus recently towards more privacy-oriented identification and authentication has resulted in several government entities pushing legislative and regulatory procedures to deal with the collection, processing, and storage of biometric and behavioural data. The most notable of these is the European Union's 2016 \gls{GDPR} \cite{Eur18}, and the \gls{CCPA} of 2018 \cite{Cal18}, both of which provide strong guarantees of privacy and security for the user's biometric data under their jurisdiction. As the user's security and privacy when interacting with a biometric system are the main focus of this research, the goal of this paper is to abstract the function of a biometric system and formalise the privacy and security requirements. We consider two approaches to security and privacy. The first is information-theoretic security and privacy, and the second is computational security and privacy.

\subsection*{Contribution.} The primary contribution of this paper is the development of a framework for ensuring the security and privacy of biometric system architectures under specified computational assumptions. In particular, we focus on the definition of an ideal functionality for a biometric authentication system that can be used to prove the security and privacy of biometric system construction through  universal composability \cite{Can01}.By leveraging the power of universal composability, we can formally prove the security and privacy of biometric system constructions, ensuring their resilience against a wide range of attacks.

\subsection*{Organisation.} The rest of this paper is organised as follows: Section \ref{sec:2.1}
provides some of the background and details the preliminaries necessary for the rest of the paper.
Section \ref{sec:2.2} introduces the reader to the biometric system setting, including high-level
abstractions of the verification and identification problems and examples of practical biometric
system constructions.

\section{Preliminaries}\label{sec:2.1}

In this section, we establish the foundational concepts and techniques that underpin our work. We start by defining metric spaces and giving some examples of metric spaces that we will use in the course of the paper. Next, we provide a high-level overview of
computational learning theory  exploring how algorithms can be trained to embed data into a metric space. Finally, we introduce neural networks as a specific type of learning algorithms, highlighting key techniques and design principles used in contemporary neural network architectures.

\subsection{Metric Spaces}

Metric spaces can be used to abstractly model biometric data, allowing us to perform comparisons
between to distinct biometrics to measure their ``closeness'', following the example of work by
Dodis \emph{et al.} \cite{DRS04}, Dodis \emph{et al.} \cite{Dod+06} and Boyen \emph{et al.}
\cite{Boy+05}. In this section, we formally define the notion of a metric space common to various
fields of mathematics and provide some examples of metric spaces.

\begin{definition}[Metric Space]
  A metric space is defined by the pair $\mathcal{M} \define (S, \Delta)$, where $S$ is a set and
  $\Delta : S \times S \rightarrow \mathbb{R}_{\geq 0}$ is a distance function satisfying the
  following properties:
  \smallskip

  \noindent 
  \hfill
  \begin{minipage}{.525\textwidth}
    \begin{equation}\label{eq:2.1}
      \forall x, y \in S : \Delta(x, y) = 0 \iff x = y
    \end{equation}
  \end{minipage}
  \hfill
  \begin{minipage}{.45\textwidth}
    \begin{equation}\label{eq:2.2}
      \forall x, y \in S : \Delta(x, y) = \Delta(y, x)
    \end{equation}
  \end{minipage}
  \hfill
  \begin{minipage}{\textwidth}
    \begin{equation}\label{eq:2.3}
      \forall x, y, z \in S : \Delta(x, z) \leq \Delta(x, y) + \Delta(y, z)
    \end{equation}
  \end{minipage}
  \hfill
  \smallskip

  \noindent where \emph{(\ref{eq:2.1})} implies both positivity and that $\forall x \in S :
  \Delta(x, x) = 0$, \emph{(\ref{eq:2.2})} is called the symmetry property, and
  \emph{(\ref{eq:2.3})} is called the triangle inequality.
\end{definition}

\noindent To simplify notation, we typically refer to the metric space by just the set rather than
the pair of the set and distance function. This notational abuse allows cleaner notations such as $x
\in \mathcal{M}$ where $\mathcal{M}$ is the metric space where the distance function $x, y \in
\mathcal{M} : \Delta(x, y) \in \mathbb{R}_{\geq 0}$ is implicit.

In this paper, we consider the following metric spaces: the Hamming metric, the Levenshtein metric,
Euclidean metric, cosine metric, and Chebyshev metric. The Hamming and Levenshtein metrics are
considered string or edit metrics, whereas the Euclidean, Chebyshev, and cosine metrics are metrics
over $n$-dimension real number spaces. 

\begin{example}[Hamming Metric]
  The Hamming metric is defined by the set $\{0,1\}^n$, which is a space of $n$-bit length strings,
  and the distance function is defined such that for $x, y \in \{0,1\}^n$
  \[
    \Delta_\mathrm{Ham}(x, y) \define \sum^n_{i=1} \mathbf{1}[x_i \neq y_i].
  \]
  Note that for the Hamming weight defined as $\mathrm{wt}(x) \define \sum^n_{i=1} \mathbf{1}[x_i
  \neq 0]$ the distance can be also be defined as $\Delta_\mathrm{Ham}(x, y) = \mathrm{wt}(x - y)$.
\end{example}

\begin{example}[Levenshtein Metric]
  The Levenshtein metric is defined by the set $\Sigma^{\leq n}$ where $\Sigma^{\leq n}$ is an
  alphabet of strings of length $\leq n$. The distance function is defined recursively such that for
  $x, y \in \Sigma^{\leq n}$ where $i = \lvert x \rvert$ and $j = \lvert y \rvert$
  \[
    \Delta_{\mathrm{Lev}, x, y} (i, j) = \begin{cases}
      \max(i, j) & \text{if}\ \min(i, j) = 0 \\
      \min \left[\begin{aligned}
        &\Delta_\mathrm{Lev}(i - 1, j) + 1, \\
        &\Delta_\mathrm{Lev}(i, j - 1) + 1, \\
        &\Delta_\mathrm{Lev}(i - 1, j - 1) + \mathbf{1}[x_i \neq y_i]
      \end{aligned} \right] & \text{otherwise}
    \end{cases}.
  \]
\end{example}

\noindent The intuition for the Levenshtein distance between two strings is the count of the
smallest number of single-character edits (operations consisting of insertion, deletion, or
substitution) required to convert one string to another. We simplify the notation for the
Levenshtein distance as $\Delta_\mathrm{Lev}(x, y)$ for convenience.

\begin{example}[Euclidean Metric]
  The Euclidean metric is defined by the set $\mathcal{M} \define \mathbb{R}^n$, which is the
  $n$-dimensional vector space over the real numbers. The magnitude of a vector $\mathbf{x} \in
  \mathbb{R}^n$ in the Euclidean metric is given by the $\ell_2$ norm $\lVert\mathbf{x}\rVert_2 =
  \sqrt{\sum^n_{i = 1} \mathbf{x}[i]^2}$ and the distance between vectors $\mathbf{x}, \mathbf{y}
  \in \mathbb{R}^n$ is $\Delta_\mathrm{Euc}(\mathbf{x}, \mathbf{y}) = \lVert \mathbf{x} - \mathbf{y}
  \rVert_2$.
\end{example}

\noindent In some cases, such as when the magnitude of vectors gives no discerning information, the
cosine similarity can be useful. The cosine similarity of two non-zero vectors in an Euclidean space
is a measure of the difference in their orientation rather than magnitude. The cosine similarity is
not technically a valid distance function as it does not obey the triangle inequality and, therefore,
cannot be used to define a metric space. For notational convenience, we still defined the cosine
similarity in the spirit of a distance function between $\mathbf{x}, \mathbf{y} \in \mathbb{R}^n$ as
\[
  \Delta_\mathrm{Cos}(\mathbf{x}, \mathbf{y}) = \frac{\langle \mathbf{x}, \mathbf{y} \rangle}{\lVert
  \mathbf{x} \rVert_2 \lVert \mathbf{y} \rVert_2}.
\]

\begin{example}[Chebyshev Metric]
  The Chebyshev metric is defined by the set $\mathbb{R}^n$ which is the $n$-dimensional vector
  space over the real numbers. The magnitude of a vector $\mathbf{x} \in \mathbb{R}^n$ in the
  Chebyshev metric is given by the $\ell_\infty$ norm $\lVert \mathbf{x} \rVert_\infty = \max^n_{i =
  1} \lvert \mathbf{x}[i] \lvert$ and the distance between vectors $\mathbf{x}, \mathbf{y} \in
  \mathbb{R}^n$ is $\Delta_\mathrm{Che}(\mathbf{x}, \mathbf{y}) = \lVert \mathbf{x} - \mathbf{y}
  \rVert_\infty$.
\end{example}

\noindent The Euclidean metric is the canonical example of a metric space for an $n$-dimensional vector
space over the reals, the intuition for which is simply the geometric distance between two vectors
in a Cartesian space. The abstraction of the Euclidean norm ($\ell_2$) to $\ell_p$ norms for vector
spaces results in the Chebyshev metric, where the Chebyshev distance is the limiting case of the
$\ell_p$ norm where $p \to \infty$.

\subsection{Learning Framework}

The biometric system constructions considered in this paper will be
heavily based on classification algorithms. Classification algorithms take input from an arbitrary space of inputs and learns how to embed that input into a representation space such that the
distance between the embeddings of correlated inputs is minimised. For example, when given a face
biometric $\mathbf{A} \in \mathbb{R}^{n \times m}$ the learning algorithm learns how to embed
$\mathbf{A}$ into the Euclidean metric $(\mathbb{R}^k, \Delta_\mathrm{Euc})$ such that vectors
representing faces with similar features have smaller distances between them. In this section, we
discuss how to construct classification algorithms using learning algorithms.

\subsubsection*{Supervised Learning.} In this paper, we will look exclusively at supervised
learning algorithms, which is the case for most biometric feature extractors based on classification
algorithms commonly encountered in the wild. The supervised learning considered in this section will
be parametrised by the input space $\mathcal{X} \subseteq \mathbb{R}^d$, output space $\mathcal{Y} =
[k]$ where $k \in \mathbb{N}$ is a set of class labels, and data distribution $\mathcal{D}$ over the
pairs $\mathcal{X} \times \mathcal{Y}$. Let $\mathcal{H} \define \{h_i : \mathcal{X} \to \mathcal{Y}
\}_{i \in [m]}$ denote a hypothesis class, which is a finite collection of functions that map the
input space to the output space.

\begin{definition}[PPT Training Algorithm]
  For a hypothesis class $\mathcal{H}$, a \emph{PPT} training algorithm $\mathsf{Train}_\mathcal{D}
  : \mathbb{N} \to \mathcal{H}$ is a randomised algorithm with sample access to a distribution
  $\mathcal{D}$ such that $\forall \lambda \in \mathbb{N}$ it runs in polynomial time in $\lambda$
  and returns $h_\lambda \in \mathcal{H}$.
\end{definition}

\[
  \mathrm{Err}_\mathcal{D}(h) \define \Pr_{(x,y) \sim \mathcal{D}}[h(x) \neq y]
\]

\subsubsection*{The PAC Learnability Framework.} Let $\mathrm{loss} : \mathcal{Y} \times \mathcal{Y} \to
\mathbb{R}_{\geq 0}$
\[
  \argmin_{h \in \mathcal{H}} \frac{1}{n} \sum^n_{i=1} \mathrm{loss}(h(x_i), y_i)
\]
\[
  \Ex_{(x, y) \sim \mathcal{D}}[\mathrm{loss}(h(x), y)]
\]

\subsection{Deep Neural Networks}

The results of the work conducted in this paper are intended for algorithms that are capable of
learning biometric representations that can be efficiently compared and matched. Given their
popularity for practical applications, and some additional beneficial properties we will outline in
this section, we focus on providing concrete results based on Feedforward Neural Networks (FNN).
FFNs are a subclass of a type of learning algorithm called a neural network. The ``modern'' variants
of neural networks have been known about in one form or another since 1943 due to work by McCulloch
and Pitts, who proposed the first binary artificial neuron \cite{MP43}. The current trend in, what
has now been dubbed, deep learning started around 2006 with work by authors such as Hinton \emph{et
al.} \cite{HOT06}, who showed that layering artificial neurons circumvented the issues that Minsky
and Papert discovered in 1969 \cite{MP69}\footnote{There was also earlier works by Ivakhnenko
\cite{Iva70,Iva71} and Amari \cite{Ama67} also were resilient to the problems proposed by Minsky and
Papert, which were lesser known due to the East-West divide in the field of neural network
research}.

The concept of FNNs was derived directly from the original works of Ivahnenko \cite{Iva70,Iva71} and
Amari \cite{Ama67}, who took the binary artificial neurons proposed by McCulloch and Pitts
\cite{MP43} and joined them together to carry out complex computations. This section will consider
two approaches to describing FNNs, each with advantages and disadvantages. First, we describe basic
FNNs using directed graphs, similar to computational circuits. By taking a graph-based approach, we
can prove the expressive power that FNNs possess, enabling them to compute arbitrary functions. We
follow this by taking the more modern approach of compactly describing an FNN by layers using linear
algebra. This approach may be familiar to those with prior experience with the style found in the
current deep-learning literature.

\begin{definition}[Feedforward Neural Network]
  A \emph{FNN} is a directed acyclic graph $G = (V,E)$ and a weight function over the edges $w : E
  \to \mathbb{R}$. A vertex in the graph is called a neuron modelled by the scalar function $\sigma
  : \mathbb{R} \to \mathbb{R}$ called an activation. Let $x_1, \ldots, x_n \in V_1$ denote the
  $n$ input layer of vertices in a $k$ layer \emph{FNN}\footnote{To simplify the definition, we
  assume that the FNN is organised into $k \in \mathbb{N}$ layers denoted by $V_{i \in [k]}$ such
  that $V = \bigcup^k_{i=1}V_i$.}. $m \leq n$
  \[
    u = \sigma \left( \sum^m_{i=1} w_i(v_i) \right)
  \]
\end{definition}

\noindent There are three classical activations that can be found in historical literature related
to FNNs which remain useful to the following discussion:
\smallskip

\noindent 
\begin{minipage}{.325\textwidth}
  \begin{equation}\label{eq:2.4}
    \sigma : x \mapsto \mathrm{sign}(x)
  \end{equation}
\end{minipage}
\hfill
\begin{minipage}{.325\textwidth}
  \begin{equation}\label{eq:2.5}
    \sigma : x \mapsto \mathbf{1}[x > 0]
  \end{equation}
\end{minipage}
\hfill
\begin{minipage}{.325\textwidth}
  \begin{equation}\label{eq:2.6}
    \sigma : x \mapsto \frac{1}{1 + e^{-x}}
  \end{equation}
\end{minipage}
\smallskip

\noindent where (\ref{eq:2.4}) is called the sign function, (\ref{eq:2.5}) is the called the
threshold function, and (\ref{eq:2.6}) is called the sigmoid function, which is a smooth
approximation of the threshold function.

\begin{definition}[Multi-Layer Perception]
  A \emph{MLP} is a \emph{FNN} where each neuron is fully connected.
\end{definition}

$\mathcal{H}_{G, \sigma, w} \define \left\{ h^{G, \sigma, w}_i : \mathbb{R}^{\lvert V_1 \rvert} \to
\mathbb{R}^{\lvert V_k \rvert} \right \}_{i \in [m]}$

\begin{lemma}\label{lem:2.1}
  Let $C_n = (V, E)$ be a Boolean circuit of depth $d \in \mathbb{N}$ that computes the function
  $f$. There exists an \emph{MLP} of depth $d$ that computes $f$.
\end{lemma}

\subsubsection*{Linear Layers.} The linear layers, or fully-connected layers, of a network are
essentially just the MLP. He \emph{et al.} \cite{He+15}.
\[
  h : \mathbf{x} \mapsto \mathbf{x}^\intercal \mathbf{W} + \mathbf{b}
\]

\subsubsection*{Activation Functions.} A classic example of an activation function is the Rectified
Linear Unit (ReLU) function, first proposed by Fukushima \cite{Fuk69,FM82}. Later, it was found that
the ReLU activation function performed particularly well during the training process of FNNs,
particularly a subclass of FNN architectures called Convolutional Neural Networks (CNN), resulting
in the widespread use of ReLU.
\[
    \sigma_i : \mathbf{x} \mapsto \frac{e^{\mathbf{x}[i]}}{1 + e^{\mathbf{x}[i]}}.
\]
By convention, we assume that activation functions, such as ReLU, are applied element-wise to an
input vector.
\[
    (\mathbf{x})^+_i : \mathbf{x} \mapsto \max(\mathbf{x}[i], 0).
\]
$\mathrm{softmax} : \mathbb{R}^n \to [0, 1]^n$
\[
    \mathrm{softmax}_i : \mathbf{x} \mapsto \frac{e^{\mathbf{x}[i]}}{\sum^n_{j=1}e^{\mathbf{x}[j]}}
\]

\subsubsection*{Convolution and Pooling Layers.} Krizhevsky \emph{et al.} \cite{KSH12} LeCun
\emph{et al.} \cite{Cun+89,Cun+98}. For a filter matrix $\mathbf{F} \in \mathbb{R}^{k \times \ell}$
and and input matrix $\mathbf{X} \in \mathbb{R}^{n \times m}$ such that $k, \ell \leq n, m$ the
convolution of the two matrices is denoted as $\mathbf{Y} = \mathbf{F} \ast \mathbf{X} \in
\mathbb{R}^{n \times m}$ where element-wise the convolution operator $\ast$ is defined as
\[
  \mathrm{conv}_{i,j} : \mathbf{X} \mapsto \sum^k_{u=1} \sum^\ell_{v=1} \mathbf{F}[u,v] \cdot
  \mathbf{X}[i - u,j - v].
\]
Max pooling Yamaguchi \emph{et al.} \cite{Yam+90} and average pooling. 
\[
  \mathrm{maxpool}_{i, j} : \mathbf{X} \mapsto \max^k_{u=1} \max^\ell_{v=1} \mathbf{X}[i, j]
\]
\[
  \mathrm{avgpool}_{i, j} : \mathbf{X} \mapsto \frac{1}{k\ell} \sum^k_{u=1} \sum^\ell_{v=1} \mathbf{X}[i, j].
\]

\subsubsection*{Gradient Descent and Backpropagation}

\[
  \mathbf{w}_{n+1} = \mathbf{w}_n - \alpha \nabla f(\mathbf{w}_n)
\]

\section{The Biometric System Setting}\label{sec:2.2}

In this section, we describe the core functionality of a biometric system and the abstract model of
biometric systems that will be useful for subsequent security and privacy analysis. A biometric system comprises several interdependent components, each playing a specific role to achieve the system’s primary objectives. To establish a comprehensive understanding of these systems prior to detailed modeling, we will review these components in depth.. A biometric system operates in one of two modes: verification (authentication) or identification. The verification and identification
operations of a biometric system will be formulated as two problems; the goal of a system is to
implement an acceptable solution to the problem. The acceptability of a solution to one of the
biometric problems is usually determined \emph{via} heuristic performance metrics, which shall also
be discussed in this section also.

\subsection{Biometric System Overview}

The canonical explanation of a traditional biometric system  is outlined in the seminal work of Jain, Ross, and Prabhakar \cite{JRP04}.  A biometric system, at its core, functions as a sophisticated pattern recognition system that processes biological or behavioral characteristics to establish or verify identity. Hence, it can be said that biometric recognition comes in two distinct modes: authentication (verification) and identification, each serving as powerful access control mechanisms when implemented appropriately.

\begin{figure}
  \centering
  \begin{tikzpicture}[
      heading/.style={font=\scshape\bfseries},
      box/.style={draw,thick,font=\scshape},
      en/.style={thick},
      ve/.style={thick,gray},
      ide/.style={thick,dashed,gray},
      key/.style={right,black,font=\scriptsize}
    ]
    \node[heading] at (0,0.25) {Sensor}; 
    \node[box,minimum width=1cm,minimum height=1cm,label={[font=\scshape]180:interface}] (I) at (0,-1) {};
    \node[box,minimum width=.9cm,minimum height=.9cm] at (0,-1) {};
    \node[circle,box,minimum width=.5cm,minimum height=.5cm,label={[font=\scshape]180:sensor}] (S) at (0,-2.5) {};
    \draw[->,en,transform canvas={xshift=-5pt}] (I) -- (S);
    \draw[->,ve] (I) -- (S);
    \draw[->,ide,transform canvas={xshift=5pt}] (I) -- (S);
    \node[circle,box,minimum width=.4cm,minimum height=.4cm] (S) at (0,-2.5) {};
    \node[heading] at (3,0.25) {Extractor}; 
    \node[box,minimum width=.75cm,minimum height=.75cm] (QC) at (2,-1) {qc};
    \node[box,minimum width=.75cm,minimum height=.75cm] (Ex) at (4,-1) {ext};
    \draw[->,en,transform canvas={yshift=5pt}] (QC) -- (I);
    \draw[->,ve] (QC) -- (I);
    \draw[->,ide,transform canvas={yshift=-5pt}] (QC) -- (I);
    \draw[->,en,transform canvas={yshift=5pt}] (QC) -- (Ex);
    \draw[->,ve] (QC) -- (Ex);
    \draw[->,ide,transform canvas={yshift=-5pt}] (QC) -- (Ex);
    \draw[->,en] ([yshift=5pt]S.east) -- ++(1.6,0) -- ([xshift=-5pt]QC.south);
    \draw[->,ve] (S.east) -| (QC.south);
    \draw[->,ide] ([yshift=-5pt]S.east) -- ++(1.95,0) -- ([xshift=5pt]QC.south);
    \node[heading] at (6,0.25) {Database}; 
    \node[box,cylinder,aspect=0.3,minimum width=1.75cm,minimum height=1cm,shape border rotate=90] (DB) at (6,-2.5) {db};
    \draw[->,en] (Ex.south) |- (DB.west);
    \node[heading] at (9,0.25) {Matcher}; 
    \node[diamond,box] (M) at (8,-1) {m};
    \draw[->,ve,gray,transform canvas={yshift=2.5pt}] (Ex) -- (M);
    \draw[->,ide,transform canvas={yshift=-2.5pt}] (Ex) -- (M);
    \draw[->,ve] ([yshift=2.5pt]DB.east) -- ++(1,0) -- ([xshift=-2.5pt]M.south);
    \draw[->,ide] ([yshift=-2.5pt]DB.east) -- ++(1.18,0) -- ([xshift=2.5pt]M.south);
    \node[box,minimum width=3.25cm] (V) at (11,-.65) {verification};
    \node[box,minimum width=3.25cm] (I) at (11,-1.4) {identification};
    \draw[->,ve] (M.east) -- (V.west);
    \draw[->,ide] (M.east) -- (I.west);
    \node[box,minimum width=2.75cm,minimum height=1.25cm] at (11.3,-2.75) {};
    \draw[->,en] (10.05,-2.35) -- (10.7,-2.35) node[key] {Enrollment};
    \draw[->,ve] (10.05,-2.75) -- (10.7,-2.75) node[key] {Verification};
    \draw[->,ide] (10.05,-3.15) -- (10.7,-3.15) node[key] {Identification};
  \end{tikzpicture}
  \caption{
    A common biometric system architecture, adapted from work by Jain \emph{et al.\@}
    \cite{JFR07}.
  }
\end{figure}
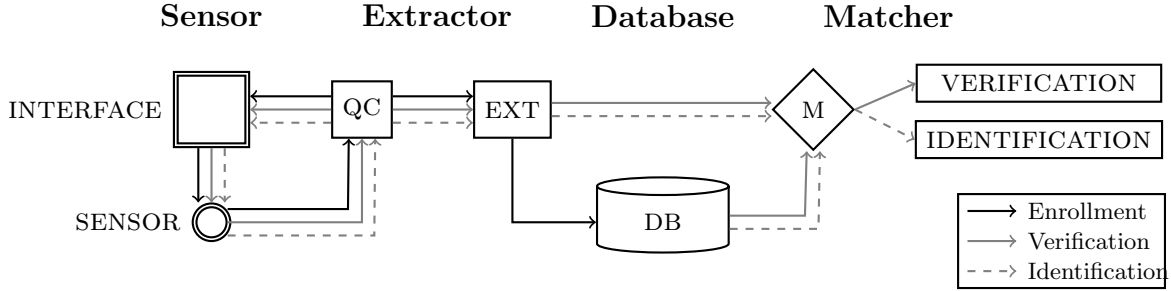

\subsubsection{Biometric System Modules.} Both the authentication and identification configuration of a biometric recognition system share four main modules. The following is a brief, high-level, overview of each of the modules that constitute and are common to all biometric systems that can be found in active use \cite{JRP04,Bha+15,ZG17,REZ20,App22}.

\begin{itemize}
  \item \emph{Sensor Module.} The first is the sensor module, which is responsible for capturing
    the biometric data from a user. Different biometrics require different types of sensors and
    there are a myriad of different sensors for each unique biometric data. Some examples of sensing
    biometric data include the recognition of faces using simple RGB cameras \cite{LJ11} or more
    complex true depth cameras used in Apple's proprietary Face ID \cite{App22}. Fingerprint
    biometric data tend to use built-in specially designed sensors which are now quite common on
    modern smartphones \cite{Yan+19}. Biometric systems utilising eye biometric data, such as the
    iris, tend to require far more invasive sensors and thus are less popular than their
    contemporaries \cite{BHF16,WH19}.

  \item \emph{Extraction Module.} Following the sampling of biometric data using the sensor module,
    the extraction module processes the raw biometric sample to extract the salient discriminatory
    features. Many traditional biometric systems use hand-crafted features designed by computer
    vision experts. A good amount of hand-crafted features for biometric systems were based on
    distributions of edges from biometric samples obtained using edge filtering methods such as
    scale-invariant feature transformation \cite{Low04} and histograms of oriented gradients
    \cite{DT05}. Other features were derived using domain transformation techniques, such as
    Gabor filtering to extract iris features \cite{Ngu+12}, Fourier transforms for facial
    recognition \cite{LYF01}, or wavelet transformations \cite{JLS04}. Principle component analysis
    has also been a popular approach to extract features by reducing the dimensionality of biometric
    samples. One highly notable example is the Eigenfaces technique for facial recognition
    \cite{TP91}. 

    The study of computer vision has shifted away from hand-crafted features in favour of using
    neural networks with many parameters for solving image recognition tasks. In fact, neural
    network architectures have been applied very successfully to a wide range of problems in both
    computer vision and natural language processing. Unlike the hand-crafted features discussed
    previously, deep neural network models provide end-to-end recognition, able to both learn the
    feature representation and perform classification or regression tasks on those features
    \cite{SW18,Min+23}. The most popular of these neural network architectures is a
    convolutional neural network, which uses several convolution layers to learn feature engineering
    without external intervention. Some notable examples of convolutional neural networks being
    applied to biometrics are FingerNet for fingerprint recognition \cite{Tan+17} and ArcFace for
    facial recognition \cite{Den+22}.

  \item \emph{Matching Module.} The next stage of the biometric recognition process is the
    matching module. This is the crux of the recognition system, where the validity of a challenge
    biometric is determined. The matching module generates a matching score based on the similarity
    of two biometric samples, an enrolled sample and a challenge sample \cite{JFR07}.
    Biometric matching is probabilistic by nature, as there will always be some uncertainty during
    the matching process. To regulate the probabilistic matching, a threshold value is used, in which
    a score that is equal to or above the threshold is seen as a match by the system
    \cite{CAK10}.

    Much like with the extraction module, there are a number of approaches to computing the matching
    score each with its own unique advantages and applicability domain. Various distance metrics
    have been used with varying degrees of effect, such as Hamming distance with binary codes for
    finger vein recognition \cite{XYY17}, Euclidean distance for face recognition in low dimension
    spaces generated by neural networks \cite{SKP15}, or cosine distance which works
    particularly well with models that use cosine loss functions \cite{Den+22}. Others have
    proposed using classification models to generate matching scores, such as the support vector
    machine for linear classification of extracted fingerprints \cite{YFJ12}.

  \item \emph{Database Module.} The final module that makes up a biometric system is the database
    which serves to store the enrolled biometric templates. The name biometric template refers to
    the stored biometric data or biometric feature data on a biometric system that is used during
    the matching process \cite{JNN08}. In the traditional setting of a biometric system, as in
    Jain \emph{et al.} \cite{JRP04}, a biometric template would consist of the extracted
    biometric feature data and a corresponding identifier unique to each enrolled entity. Before a
    user can be verified or identified by a biometric system they need to either enrol themselves or
    be enrolled onto the system.

   Overall, in some cases, the storing of the full user
    biometric data is unnecessary for a secure biometric system to function
    \cite{DRS04,IW09,Wan+12,Li+17}, rather ``helper'' data can be provided to achieve a
    successful match. Furthermore, databases themselves aren't a requirement for a biometric system.
    Depending on the application, a template may be stored on a smaller device \emph{i.e.}
    identification cards for access control systems that use biometric verification \cite{Noo00} or
    the increasingly popular security keys for passwordless or two-factor authentication
    \cite{Bar+21}. The fact that a database may not necessarily be used in a biometric system makes
    the name ``database module'' a misnomer in some cases; however, the name is common in the
    literature \cite{JBP96,JRP04,Phi+00,PPJ03,IW09} and will be used in this paper.
\end{itemize}

\subsection{Accuracy and Correctness}

Despite commonalities between their constituent components, biometric verification and
identification are different tasks and, therefore, require different performance statistics
\cite{Phi+00}. We will more formally define the verification and identification problems shortly,
but for now, the intuitive definitions of the verification and identification configurations of a
biometric system is given as follows: In a verification setting, an authenticating entity makes a
claim against a specified pre-enrolled identity and supplies a challenge biometric which is either
accepted or rejected by the verifier \cite{JBP96,JRP04,Bol+04}. In the identification setting an
entity only provides a challenge biometric in which the identifier attempts to find a valid identity
associated with that biometric, if one exists \cite{JRP04,JRP06,JFR07}.

In a verification configuration, a biometric system can make two types of errors: (1) a \gls{FM}
where biometric measures from two unique entities are mistaken for being from the same entity and
(2) a \gls{FNM} were biometric measures from one entity are mistaken for being from two unique
entities. A \gls{FM} and a \gls{FNM} are sometimes referred to as a false accept or false reject in
other literature \cite{JRP04,GER13}. In a perfect verification system, both the FM and FNM would be
zero; unfortunately, due to its probabilistic nature, a biometric system cannot be perfect
\cite{Phi+00}. We define the \gls{FMR} and \gls{FNMR} below.

\begin{definition}[False Match Rate]
  Let $S = \{s_i\}^n_{i=1}$ be a set of possible results. For a threshold value $t$ the
  \emph{\gls{FMR}}
  for $S$ is
  \[
    \mathrm{FMR}_t(S) = \frac{1}{n}\sum^{n}_{i=1}\mathbf{1}_{\{s_i > t\}}.
  \]
\end{definition}

\begin{definition}[False Non-Match Rate]
  Let $S = \{s_i\}^n_{i=1}$ be a set of possible results. For a threshold value $t$ the
  \emph{\gls{FNMR}}
  for $S$ is
  \[
    \mathrm{FNMR}_t(S) = \frac{1}{n}\sum^{n}_{i=1}\mathbf{1}_{\{s_i \leq t\}}.
  \]
\end{definition}

\noindent Construction of a successful biometric system in a verification configuration must
consider the balance of the \gls{FMR} and \gls{FNMR} to achieve sufficient security and accuracy. One
common method to find the correct balance is to use the \gls{ROC} of the system. The \gls{ROC} lets
us see the performance of the system across all threshold values, called operating points
\cite{JLS04}. We call the operating point at which the \gls{FMR} and \gls{FNMR} rates are the same
is called the \gls{EER}, which we define below.

\begin{definition}[Equal Error Rate]
  Let $S = \{s_i\}^n_{i=1}$ be a set of possible results and $T = \{t_i\}^m_{i=1}$ a set of
  potential threshold values. Let the threshold value which the \emph{\gls{FMR}} is the closest to
  \emph{\gls{FNMR}} on $S$ be $t = \mathrm{arg}\min^m_{i = 1} \left\lvert \mathrm{FMR}_{t_i}(S) -
  \mathrm{FNMR}_{t_i}(S) \right\rvert$. The \emph{\gls{EER}} for $S$ with possible thresholds $T$ is
  \[
    \mathrm{EER}_T(S) = \frac{\mathrm{FMR}_{t}(S) + \mathrm{FNMR}_{t}(S)}{2}.
  \]
\end{definition}

\noindent Unlike the verification configuration, the identification configuration's main performance
measure is how well it can identify a challenge biometric's true owner \cite{Phi+00}. The
performance of the system in identification configuration can be measured using the system accuracy
in a verification mode under some additional simplifying assumptions \cite{JRP04}. In essence, by
assuming that one match attempt is made per enrolled biometric template on the system and error
scores between different templates are independent; the \gls{FM} and \gls{FNM} for an identification
can be extrapolated resulting in a reasonable approximation \cite{JFR07}. Let $\mathrm{FNMR}^n$ and
$\mathrm{FMR}^n$ denote the FNMR and FMR for $n$ identities in a system database module
respectively. The approximations of FNMR and FMR for a threshold $t$ and set of outcomes $S$
\[
  \mathrm{FNMR}^n_t(S) \approx \mathrm{FNMR}_t(S) \qquad \mathrm{FMR}^n_t(S) \approx n \cdot
  \mathrm{FMR}_t(S)
\]
are sufficient when $n \cdot \mathrm{FMR} < \frac{1}{10}$. For a database with classified and
indexed templates, a more accurate performance measure can be taken by also considering error rates
of incorrect retrievals \cite{JRP04}.

\section{Feature Extraction with Representation Learning}

The machinery that makes a biometric system function is the extraction module built from learning
algorithms. This section will examine the common principles behind the learning algorithms used for
constructing the feature extraction module. In particular, we examine how algorithms can be
constructed to learn how to place biometric data into a metric space suitable for the matching
module to achieve a sufficient \gls{EER}. We begin this section by examining the framework of
learning problems, including \gls{ERM}, the \gls{PAC} framework, and the use of feature
extractors\footnote{A feature extractor, as defined in statistical or machine learning research,
serves a different purpose to the feature extraction module described as part of a biometric
system.} to improve performance and learnability. We then describe some practical constructions of
biometric feature extractor modules that we will use throughout the rest of this paper, including
fingerprint features with Gabor Filters, face features with \gls{PCA} and deep learning, and iris
features with deep learning.

\begin{remark}
  Before we begin our closer examination of the feature extraction module, let us address the lack
  of in-depth discussion of the sensor model. The goal of this paper is to abstractly model the
  biometric system for use in a privacy-preserving and secure biometric identification or
  authentication framework. Since our focus is specifically on smartphone settings, the
  difficulty involved in the discussion of the biometric sensor modules is the variety of sensors
  that are used across the myriad of smartphones available on the market. For example, Apple's Face
  ID \cite{App22} and Touch ID \cite{App23} use sensors that may differ significantly from the
  Google Pixel optical fingerprint sensors \cite{Goo24}. Furthermore, due to the proprietary nature
  of some hardware products, it can be difficult to determine exactly how they function. Therefore,
  we limit our discussion to the process of learning useful representations from,
  what are assumed to be, quality biometric data.
\end{remark}





\section{Biometric Verification}

\subsection{Biometric Verification Formulation}

Using the components defined above, we can define the two configurations of a biometric system,
starting with the verification configuration. In verification configuration, a biometric system
validates a user's identity by comparing challenge biometric data with a biometric template known
\emph{a priori} \cite{JRP04}. With the previous preliminaries in mind, we formally define the
biometric verification problem in the style of a computational decision problem. Computational
decision type problems involve a function $f : \{0,1\}^\ast \to \{0,1\}$ which solves the decision
problem of $S \subseteq \{0,1\}^\ast$ if for every $x \in S$ it holds that $f(x) = 1$.

\begin{definition}[The Biometric Verification Problem]
  Let $(\mathcal{M}, \Delta)$ define a metric space and $g : \mathcal{X} \to \mathcal{M}$ be a
  mapping for the set of biometric source data $\mathcal{X}$ to that metric space. Given the
  functions $\ell : \mathbb{N} \to \mathbb{N}$ and $t : \mathbb{N} \to \mathbb{R}_+$, the biometric
  verification problem is defined by
  \[
    \Pr \left[ 
      \Delta(g(x'), x) \leq t(\lambda)
      \middle|
      \begin{aligned}
        & \mathcal{D} := \{(\mathsf{id}_i, g(x_i))\}^{\ell(\lambda)}_{i=1} \\
        & \text{for}\ \mathsf{id}_i \samp \{0,1\}^{\leq \lambda}\ \text{and}\ x_i \samp \mathcal{X}
        \\
        & \text{choose}\ \mathsf{id}' \in \{0, 1\}^{\leq \lambda} \\
        & \text{find}\ x\ \text{where}\ (\mathsf{id}', x) \in \mathcal{D} \\
        & x' \samp \mathcal{X} \\
      \end{aligned}
    \right] \geq 1 - \varepsilon(\lambda)
  \]
  where the equality must hold for sufficiently large $\lambda \in \mathbb{N}$ such that each 
  $\mathsf{id}_i$ is unique with high probability and a negligible function $\varepsilon :
  \mathbb{N} \to [0, 1]$.
\end{definition}

\noindent In the definition above, the functions $\ell$ and $t$ allow the success threshold and size
of the database to be functions of $\lambda$; effectively parametrising the biometric verification
problem by $\lambda$. We now define the construction of a biometric verification system that covers
some of the details that are somewhat glossed over in the biometric verification problem definition.

\begin{construction}[Biometric Verification System]\label{cons:2.1}
  We define a biometric verification system as an interactive protocol between a user $\mathcal{P}$
  proving their knowledge of an enrolled template and a verifier $\mathcal{V}$. Let $(\mathcal{M},
  \Delta)$ be a metric space and $\mathcal{X}$ be the space of input biometric data. A biometric
  verification system consists of three subroutines or methods $\mathsf{init}$, $\mathsf{enroll}$,
  and $\mathsf{verify}$ defined as follows:
  \begin{itemize}
    \item $\mathsf{init}(\lambda) \to \mathsf{pp}$: In this subroutine, the following public and
      private parameters are defined by $\mathcal{V}$:
      \begin{itemize}
        \item Define the mapping $g : \mathcal{X} \to \mathcal{M}$ such that $g = h \circ f$ where
          $h : \mathcal{F} \to \mathcal{M}$ for $h$ in the hypothesis class $\mathcal{H}$ and $f :
          \mathcal{X} \to \mathcal{F}$ is a feature extractor. Assume $h$ is the \gls{PAC}
          hypothesis for mapping the feature space $\mathcal{F}$ to the metric space $\mathcal{M}$.
        \item $\mathcal{V}$ then defines $\mathcal{D} \subseteq \{0, 1\}^{\leq \lambda} \times
          \mathcal{M}$ for storage of biometric templates such that $\lvert \mathcal{D} \rvert \leq 
          \ell(\lambda)$ where $\ell : \mathbb{N} \to \mathbb{N}$.
        \item EER.
      \end{itemize}
      $\mathcal{V}$ returns the tuple of public parameters $\mathsf{pp} := (\lambda, g)$ to use in the
      subsequent subroutines. For simplicity, assume $\mathsf{init}$ is only executed once per
      biometric system in verification configuration.
    \item $\mathsf{enroll}_\mathsf{pp}(x) \to \mathsf{id}$: The enrollment subroutine begins by
      $\mathcal{P}$ sampling $x \samp \mathcal{X}$ and sending $g(x)$ to $\mathcal{V}$. The
      biometric data $x$ is called the enrolled biometric. $\mathcal{V}$ generates the identifier
      $\mathsf{id} \samp \{0, 1\}^{\leq \lambda}$ and performs $\mathcal{D} := \mathcal{D} \cup
      \{(\mathsf{id}, g(x))\}$, storing the biometric data in the database. $\mathcal{V}$ can then
      directly send $\mathsf{id}$ to $\mathcal{P}$, or it can publish it depending on the intended
      use of the system.
    \item $\mathsf{verify}_\mathsf{pp}(\mathsf{id}, x) \to b$: $\mathcal{V}$ returns $b :=
      \mathbf{1}_{\{\Delta(g(x), y) \leq t(\lambda)\}}$.
  \end{itemize}
  All randomness in the protocol is taken over the internal coins of $\mathcal{P}$ and
  $\mathcal{V}$.
\end{construction}

\begin{figure}[ht]
  \centering
  \begin{tikzpicture}[
      node/.style={circle,draw,thick,minimum width=1cm,font=\scshape}
    ]
    \node[node] (U) {U}; 
    \node[node,right=2cm of U] (D) {D}; 
    \node[node,right=2cm of D] (A) {V}; 
    \node[node,above right=2cm of A] (DB) {DB}; 
    \node[node,below right=2cm of A] (M) {M}; 
    \draw[->,thick] (U) -- node[above] {$w'$, $\mathsf{id}$} (D);
    \draw[->,thick] (D) -- node[above] {$h(w')$, $\mathsf{id}$} (A);
    \draw[-left to,thick,transform canvas={xshift=1pt,yshift=1pt}] (A) -- node[above,rotate=-45] {$b, b'$} (M);
    \draw[left to-,thick,transform canvas={xshift=-1pt,yshift=-1pt}] (A) -- node[below,rotate=-45] {$\Delta(b, b')$} (M);
    \draw[-left to,thick,transform canvas={xshift=-1pt,yshift=1pt}] (A) -- node[above,rotate=45] {$b', \mathsf{id}$} (DB);
    \draw[left to-,thick,transform canvas={xshift=1pt,yshift=-1pt}] (A) -- node[below,rotate=45] {$b$} (DB);
  \end{tikzpicture}
  \caption{
    A biometric verification system model with indications of the generic information flows based on
    work by Jain \emph{et al.\@} \cite{JFR07} and Construction \ref{cons:2.1}. In the above figure, we
    assume that the $\mathsf{init}$ method of Construction \ref{cons:2.1} has already been run.
  }
  \label{fig:2.2}
\end{figure}
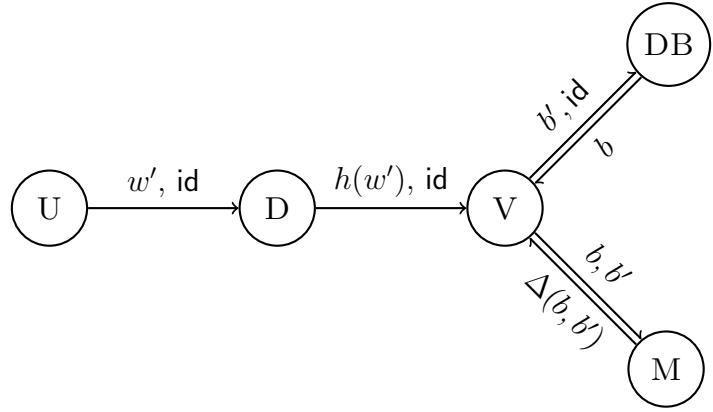

\noindent Figure \ref{fig:2.2} visually illustrates the construction above. We will now look at two
concrete examples of biometric verification systems that have been used in practical applications of
biometric verification. The first is a verification system using fingerprint features in the Hamming
metric, and the second is a verification system that uses face features in the Euclidean metric.

\begin{lemma}
  For a metric space $(\mathcal{M}, \Delta)$ and mapping $h : \mathcal{X} \to \mathcal{M}$ the
  Construction \ref{cons:2.1} is a solution to the biometric verification problem.
\end{lemma}

\begin{example}[Fingerprint Verification in the Hamming Metric]
  Let the metric space be defined by $(\{0,1\}^n, \Delta_\mathrm{Ham})$
\end{example}


\section{Biometric Identification}

\subsection{Biometric Identification Formulation}

Computational search type problems involve a function $f : \{0, 1\}^\ast \to \{0, 1\}^\ast \cup
\{\bot\}$ solves the search problem for the relationship $R \subseteq \{0, 1\}^\ast \times \{0,
1\}^\ast$ if for every $x \in \{0, 1\}^\ast$ it holds that $f(x) \in R(x)$ where $R(x) := \{y \mid
(x, y) \in R\}$ otherwise $f(x) = \bot$.

The identification configuration of a biometric system attempts to recognise a challenge biometric
by searching the database of all enrolled templates. Unlike the one-to-one comparison of the
verification configuration, identification performs a one-to-many comparison to establish the
identity of an individual \cite{JRP04}. Identification is the key component in a negative
recognition system, where a system verifies that a subject is who they deny they are, preventing
them from claiming multiple identities \cite{Way01}. While verification is possible through
several authentication mechanisms, negative recognition is unique to biometrics identification. The
biometric identification problem is formally posited in the following definition.

\begin{definition}\label{def:biometric-identification}
    Given an $n$-dimensional biometric input $\mathbf{x} \in \mathbb{R}^n$ the biometric identification
    problem is to determine which identity, if any, a $m$-dimensional feature vector $f(\mathbf{x})
    \in \mathbb{R}^m$ where $m < n$ corresponds to out of the set $\{(\mathsf{ID}_i,
    \mathbf{x}'_i)\}^N_{i=1}$ of enrolled biometric templates.
\end{definition}

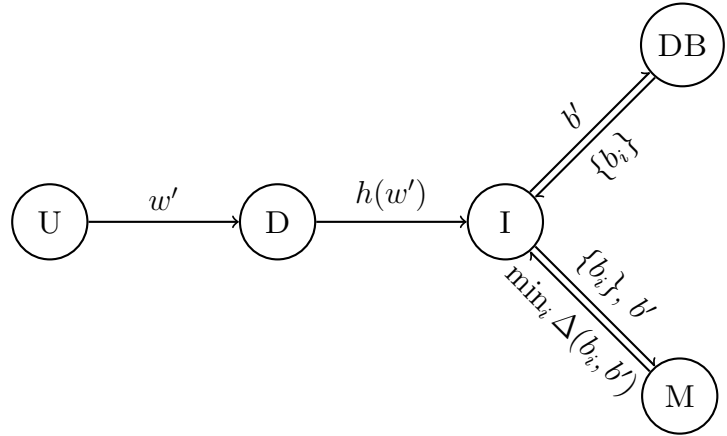
\begin{figure}[ht]
  \centering
  \begin{tikzpicture}[
      node/.style={circle,draw,thick,minimum width=1cm,font=\scshape}
    ]
    \node[node] (U) {U}; 
    \node[node,right=2cm of U] (D) {D}; 
    \node[node,right=2cm of D] (A) {I}; 
    \node[node,above right=2.25cm of A] (DB) {DB}; 
    \node[node,below right=2.25cm of A] (M) {M}; 
    \draw[->,thick] (U) -- node[above] {$w'$} (D);
    \draw[->,thick] (D) -- node[above] {$h(w')$} (A);
    \draw[-left to,thick,transform canvas={xshift=1pt,yshift=1pt}] (A) -- node[above,rotate=-45]
    {$\{b_i\}$, $b'$} (M);
    \draw[left to-,thick,transform canvas={xshift=-1pt,yshift=-1pt}] (A) -- node[below,rotate=-45]
    {$\min_{i} \Delta(b_i, b')$} (M);
    \draw[-left to,thick,transform canvas={xshift=-1pt,yshift=1pt}] (A) -- node[above,rotate=45] {$b'$} (DB);
    \draw[left to-,thick,transform canvas={xshift=1pt,yshift=-1pt}] (A) -- node[below,rotate=45]
    {$\{b_i\}$} (DB);
  \end{tikzpicture}
  \caption{
    A biometric identification system model with indications of the generic information flows based
    on work by Jain \emph{et al.\@} \cite{JFR07} and Construction \ref{cons:2.2}. In the above figure,
    we assume that the $\mathsf{init}$ method of Construction \ref{cons:2.2} has already been run.
  }
  \label{fig:2.3}
\end{figure}

Much like the Definition of the biometric verification problem, Definition
\ref{def:biometric-identification} encapsulates the identification problem as solved by a biometric
system in the identification configuration. However, unlike in verification, the task is to take only
a given biometric data $\mathbf{x} \in \mathbb{R}^n$ and match the feature vector $f(\mathbf{x}) \in
\mathbb{R}^m$ to an enrolled template $(\mathsf{ID}, \mathbf{x}')$. The matcher module of the system
requires some slight modification to accomplish this; rather than output the binary value $\{\top,
\bot\}$, the matcher searches through a set of all enrolled templates and returns the $\mathsf{ID}$ of
the highest value match that exceeds the threshold. More formally, the matcher performs
\[
    \mathsf{ID} = \begin{cases}
        \mathsf{ID}_i & \text{if}\ \max_{i=1}^N \mathsf{dist}(f(\mathbf{x}_i), f(\mathbf{x}')) \geq t\\
        \bot & \text{otherwise}
    \end{cases}
\]
where $t \in \mathbb{R}$ is the threshold defined by the system, $\mathsf{dist} : \mathbb{R}^m
\times \mathbb{R}^m \rightarrow \mathbb{R}$ is an arbitrary distance function between two
$m$-dimensional feature vectors, $N$ is the number of enrolled biometric templates and the output
of the biometric verification is $\{\mathsf{ID}_1, \ldots, \mathsf{ID}_N\} \cup \{\bot\}$. A diagram
illustrating the basic functionality of enrolment and identification of a biometric identification
system can be seen in Figure \ref{fig:2.3}.

\begin{definition}[The Biometric Identification Problem]
  Let $(\mathcal{M}, \Delta)$ define a metric space and $g : \mathcal{X} \to \mathcal{M}$ be a
  mapping for the set of biometric source data $\mathcal{X}$ to that metric space. Given the
  functions $\ell : \mathbb{N} \to \mathbb{N}$ and $t : \mathbb{N} \to \mathbb{R}_+$, the biometric
  verification problem is defined by
  \[
    \Pr \left[ 
      \Delta(g(x'), x) \leq t(\lambda)
      \middle|
      \begin{aligned}
        & \mathcal{D} := \{(\mathsf{id}_i, g(x_i))\}^{\ell(\lambda)}_{i=1} \\
        & \text{for}\ \mathsf{id}_i \samp \{0,1\}^{\leq \lambda}\ \text{and}\ x_i \samp \mathcal{X}
        \\
        & \text{choose}\ \mathsf{id}' \in \{0, 1\}^{\leq \lambda} \\
        & \text{find}\ x\ \text{where}\ (\mathsf{id}', x) \in \mathcal{D} \\
        & x' \samp \mathcal{X} \\
      \end{aligned}
    \right] \geq 1 - \varepsilon(\lambda)
  \]
  where the equality must hold for sufficiently large $\lambda \in \mathbb{N}$ such that each 
  $\mathsf{id}_i$ is unique with high probability and a negligible function $\varepsilon :
  \mathbb{N} \to [0, 1]$.
\end{definition}

\noindent In the definition above, the functions $\ell$ and $t$ allow the success threshold and size
of the database to be functions of $\lambda$; effectively parametrising the biometric verification
problem by $\lambda$. We now define the construction of a biometric verification system that covers
some of the details that are somewhat glossed over in the biometric verification problem definition.

\begin{construction}[Biometric Identification System]\label{cons:2.2}
  We define a biometric verification system as an interactive protocol between a user $\mathcal{P}$
  proving their knowledge of an enrolled template and a verifier $\mathcal{V}$. Let $(\mathcal{M},
  \Delta)$ be a metric space and $\mathcal{X}$ be the space of input biometric data. A biometric
  verification system consists of three subroutines or methods $\mathsf{init}$, $\mathsf{enroll}$,
  and $\mathsf{verify}$ defined as follows:
  \begin{itemize}
    \item $\mathsf{init}(\lambda) \to \mathsf{pp}$: In this subroutine, the following public and
      private parameters are defined by $\mathcal{V}$:
      \begin{itemize}
        \item Define the mapping $g : \mathcal{X} \to \mathcal{M}$ such that $g = h \circ f$ where
          $h : \mathcal{F} \to \mathcal{M}$ for $h$ in the hypothesis class $\mathcal{H}$ and $f :
          \mathcal{X} \to \mathcal{F}$ is a feature extractor. Assume $h$ is the \gls{PAC}
          hypothesis for mapping the feature space $\mathcal{F}$ to the metric space $\mathcal{M}$.
        \item $\mathcal{V}$ then defines $\mathcal{D} \subseteq \{0, 1\}^{\leq \lambda} \times
          \mathcal{M}$ for storage of biometric templates such that $\lvert \mathcal{D} \rvert \leq 
          \ell(\lambda)$ where $\ell : \mathbb{N} \to \mathbb{N}$.
        \item EER.
      \end{itemize}
      $\mathcal{V}$ returns the tuple of public parameters $\mathsf{pp} := (\lambda, g)$ to use in the
      subsequent subroutines. For simplicity, assume $\mathsf{init}$ is only executed once per
      biometric system in verification configuration.
    \item $\mathsf{enroll}_\mathsf{pp}(x) \to \mathsf{id}$: The enrollment subroutine begins by
      $\mathcal{P}$ sampling $x \samp \mathcal{X}$ and sending $g(x)$ to $\mathcal{V}$. The
      biometric data $x$ is called the enrolled biometric. $\mathcal{V}$ generates the identifier
      $\mathsf{id} \samp \{0, 1\}^{\leq \lambda}$ and performs $\mathcal{D} := \mathcal{D} \cup
      \{(\mathsf{id}, g(x))\}$, storing the biometric data in the database. $\mathcal{V}$ can then
      directly send $\mathsf{id}$ to $\mathcal{P}$, or it can publish it depending on the intended
      use of the system.
    \item $\mathsf{verify}_\mathsf{pp}(\mathsf{id}, x) \to b$: $\mathcal{V}$ returns $b :=
      \mathbf{1}_{\{\Delta(g(x), y) \leq t(\lambda)\}}$.
  \end{itemize}
  All randomness in the protocol is taken over the internal coins of $\mathcal{P}$ and
  $\mathcal{V}$.
\end{construction}

\begin{lemma}
  For a metric space $(\mathcal{M}, \Delta)$ and mapping $h : \mathcal{X} \to \mathcal{M}$ the
  Construction \ref{cons:2.2} is a solution to the biometric identification problem.
\end{lemma}

\begin{example}[Face Verification in the Euclidean Metric]
  Let the metric space be defined by $(\mathbb{R}^n, \Delta_\mathrm{Euc})$
\end{example}

\section{Conclusions}
This paper has provided a framework for the security and privacy of biometric system constructions under defined computational assumptions. It uses the concept of an ideal functionality for a biometric authentication system that can be used to prove the security and privacy of biometric system construction with a key focus on universal composability. We believe this framework can be effective in building efficient construction of biometric systems with various cryptographic primitives.

\bibliographystyle{IEEEtran}
\bibliography{main}

\end{document}

%% file: gloss.tex

\newacronym{DPT}{DPT}{Deterministic Polynomial-Time}
\newacronym{PPT}{PPT}{Probabilistic Polynomial-Time}
\newacronym{EPT}{EPT}{Expected Polynomial-Time}
\newacronym{PRF}{PRF}{Pseudo-Random Function}
\newacronym{CRHF}{CRHF}{Collision Resistant Hash Function}
\newacronym{MAC}{MAC}{Message Authentication Code}
\newacronym{PKE}{PKE}{Public Key Encryption}
\newacronym{IND-CPA}{IND-CPA}{Indistinguishable under Chosen Plaintext Attacks}
\newacronym{IND-CCA}{IND-CCA}{Indistinguishable under Chosen Ciphertext Attacks}
\newacronym{DSS}{DSS}{Digital Signature Schemes}
\newacronym{SUF-CMA}{SUF-CMA}{Strong Unforgeability under Chosen Message Attack}
\newacronym{DAG}{DAG}{Directed Acyclic Graph}

\newacronym{GDPR}{GDPR}{General Data Protection Regulation}
\newacronym{CCPA}{CCPA}{California Consumer Privacy Act}
\newacronym{HILL}{HILL}{Håstad-Impagliazzo-Levin-Luby Entropy}
\newacronym{FNM}{FNM}{False-Non Match}
\newacronym{FM}{FM}{False Match}
\newacronym{FNMR}{FNMR}{False-Non Match Rate}
\newacronym{FMR}{FMR}{False Match Rate}
\newacronym{ROC}{ROC}{Receiver Operating Characteristic}
\newacronym{EER}{EER}{Equal Error Rate}
\newacronym{ERM}{ERM}{Empirical Risk Minimisation}
\newacronym{PAC}{PAC}{Probably Approximately Correct}
\newacronym{PCA}{PCA}{Principle Component Analysis}

\newacronym{HE}{HE}{Homomorphic Encryption}
\newacronym{SWHE}{SWHE}{Somewhat-Homomorphic Encryption}
\newacronym{FHE}{FHE}{Fully Homomorphic Encryption}
\newacronym{FE}{FE}{Fuzzy Extractor}
\newacronym{SS}{SS}{Secure Sketch}
\newacronym{CFE}{CFE}{Computational Fuzzy Extractor}

\newacronym{HA}{HA}{Homomorphic Authenticator}
\newacronym{NIZK}{NIZK}{Non-Interactive Zero-Knowledge}
\newacronym{STARK}{STARK}{Scalable Transcript Argument of Knowledge}
\newacronym{SNARK}{SNARK}{Succinct Non-Interactive Argument of Knowledge}
\newacronym{CRS}{CRS}{Common Reference String}
\newacronym{UF-CMA}{UF-CMA}{Unforgeability under Chosen Message Attack}